\documentclass[11pt]{article}
\usepackage{graphicx}
\usepackage{amssymb}
\usepackage{amsmath}
\usepackage{enumitem}
\usepackage{xspace}
\usepackage[left]{lineno}
\usepackage{subfigure}
\usepackage{booktabs}
\usepackage{multirow}
\usepackage[title]{appendix}
\usepackage{float}
\usepackage{relsize}
\usepackage{hyperref}
\hypersetup{colorlinks=true,urlcolor=blue,citecolor=blue}
\usepackage{alphalph}

\newcommand{\kdpi}{$K^+ \to \pi^+ \pi^0$ }
\newcommand{\keqn}{$K^+ \to \pi^0 \pi^0 e^+ \nu$ }
\newcommand{\ketg}{$K^+ \to \pi^0 e^+ \nu \gamma$ }
\newcommand{\ket}{$K^+ \to \pi^0 e^+ \nu (\gamma)$ }

\begin{document}

\begin{titlepage}

\LARGE EUROPEAN ORGANIZATION FOR NUCLEAR RESEARCH

\vspace{15mm}

{\flushright{
\normalsize
CERN-EP-2023-069 \\
21 April 2023 \\
-- \\
Revised version: \\
22 August 2023 \\
}}

\vspace{15mm}

\begin{center}
   \textbf{\boldmath A study of the \ketg decay} 
\end{center}

\begin{center}
\begin{NoHyper}
\renewcommand{\thefootnote}{\fnsymbol{footnote}}
The NA62 Collaboration
\footnote[1]{ email: na62eb@cern.ch \\ Corresponding authors: F. Brizioli (francesco.brizioli@cern.ch), D. Madigozhin (dmitry.madigozhin@cern.ch) }
\end{NoHyper}
\end{center}

\vspace{5mm}

\abstract{
A sample of $1.3 \times 10^5$ \ketg candidates with less than 1\% background was collected by the NA62 experiment at the CERN SPS in 2017--2018. Branching fraction measurements are obtained at percent relative precision in three restricted kinematic regions, improving on existing results by a factor larger than two.  An asymmetry, possibly related to T-violation, is investigated with no evidence observed within the achieved precision.
}

\vspace{20mm}

\begin{center}
\emph{Accepted for publication in JHEP}    
\end{center}

\end{titlepage}

\newpage
\clearpage
\pagenumbering{arabic}


\section{Introduction}
A study of the \ketg decay allows a precision test of Chiral Perturbation Theory (ChPT), an effective field theory of 
Quantum Chromodynamics (QCD)
at low energies based on the chiral symmetry properties of the QCD Lagrangian~\cite{ChPT1, ChPT2, ChPT3}. In this framework, the \ketg decay is described by inner bremsstrahlung (IB) and structure dependent (SD) processes and their interference.  
Calculations of the branching fraction for the \ketg decay at different orders of approximation of radiative effects are given in~\cite{Bijnens:1992en, Braguta_PhysRevD.65.054038, Kubis:2006nh, Khriplovich:2010rz}.

The radiative decay \ketg ($K_{e3\gamma}$) is defined as a subset of the inclusive decays \ket ($K_{e3}$)
that requires a minimum photon energy ($E_\gamma$) and a range of angles between the positron and the radiative photon ($\theta_{e \gamma}$) in the kaon rest frame.
These conditions are designed to prevent the decay amplitude from diverging in the infrared ($E_\gamma \to 0$) and collinear ($\theta_{e \gamma} \to 0$) limits.

The ratio of the branching fractions of the radiative decay $K_{e3\gamma}$ to the inclusive decay $K_{e3}$ is expressed as:
\begin{equation}
    R_j =
\frac{\mathcal{B}(K_{e3\gamma^j})}{\mathcal{B}(K_{e3})} =
\frac{\mathcal{B}(K^{+} \rightarrow \pi^{0} e^{+} \nu \gamma \; | \; E_\gamma^j, \; \theta_{e \gamma}^j)}{\mathcal{B}(K^{+} \rightarrow \pi^{0} e^{+} \nu (\gamma))} ,
\label{eq:br_def}
\end{equation}
where ($E_\gamma^j$, $\theta_{e \gamma}^j$)
are the conditions corresponding to the kinematic regions labeled by the index $j$.
The definitions of the three kinematic regions used in this analysis are given in Table~\ref{tab:ke3g_stateofart}, together with theoretical~\cite{Kubis:2006nh} and experimental~\cite{Akimenko:2007zz, Polyarush:2020ocu} $R_j$ results.
Another theoretical calculation~\cite{Khriplovich:2010rz} predicts the branching fraction
$\mathcal{B}(K^{+} \rightarrow \pi^{0} e^{+} \nu \gamma \; | \; E_\gamma > 30$ MeV, $\theta_{e \gamma} > 20^\circ) = (2.72 \pm 0.10) \times 10^{-4}$,
that is converted to $R_2$ using the $K_{e3}$ branching fraction~\cite{pdg}: $R_2 = (0.54 \pm 0.02) \times 10^{-2}$.
 
\begin{table}[h]
\caption{
Definition of kinematic regions, $R_j$ expectations from ChPT $\mathcal{O}(p^6)$ calculations~\cite{Kubis:2006nh} and measurements from ISTRA+~\cite{Akimenko:2007zz} and OKA~\cite{Polyarush:2020ocu} experiments, with statistical and systematic uncertainties quoted separately.
}
    \centering
    \scriptsize
\begin{center}
\begin{tabular}{|l||c|c|c|c|}
\hline
 &  $E_\gamma^j$, $\theta_{e \gamma}^j$ & ChPT & ISTRA+ & OKA \\ \hline \hline
$R_1 \times 10^2$ & $ E_\gamma > 10$ MeV, $ \theta_{e \gamma} > 10^\circ $ & $ 1.804 \pm 0.021$ & $ 1.81 \pm 0.03 \pm 0.07$ & $ 1.990 \pm 0.017 \pm 0.021$\\ 
$R_2  \times 10^2$ & $ E_\gamma > 30$ MeV, $ \theta_{e \gamma} > 20^\circ $  & $ 0.640 \pm 0.008 $ & $ 0.63 \pm 0.02 \pm 0.03$ & $ 0.587 \pm 0.010 \pm 0.015$ \\
$R_3  \times 10^2$ & $ E_\gamma > 10 $ MeV, $ 0.6 < \cos \theta_{e \gamma} < 0.9 $  & $ 0.559 \pm 0.006$ & $ 0.47 \pm 0.02 \pm 0.03$ & $ 0.532 \pm 0.010 \pm 0.012$ \\ \hline
\end{tabular}
\end{center}
\label{tab:ke3g_stateofart}
\end{table}

The amplitude of the $K_{e3\gamma}$ decay is sensitive to T-violating contributions, which can be studied with the dimensionless T-odd observable $\xi$ and the corresponding T-asymmetry $A_\xi$, defined as:
\begin{equation}
\xi = \frac{\Vec{p_\gamma} \cdot (\Vec{p_e} \times \Vec{p_\pi})}{(M_K \cdot c)^3}, \; A_\xi = \frac{N_+ - N_-}{N_+ + N_-},
\label{eq:def_xi}
\end{equation}
where $\Vec{p}$ is the three-momentum of each particle in the kaon rest frame, $M_K$ is the charged kaon mass~\cite{pdg}, $c$ is the speed of light and $N_+ \; (N_-)$ is the number of events with positive (negative) values of $\xi$.

Most theoretical calculations of $A_\xi$, both within the Standard Model (SM)~\cite{Braguta_PhysRevD.65.054038,Khriplovich:2010rz,Muller:2006gu} and beyond~\cite{Muller:2006gu,Braguta_beyond}, predict values in the range $[-10^{-4} , -10^{-5}]$ with the exception of one SM extension~\cite{Muller:2006gu}, which quotes a value of $-2.5 \times 10^{-3}$. 
Non-zero values of $A_\xi$ in the SM originate from one-loop electromagnetic corrections.
The uncertainty of the most precise published $A_\xi$ measurement~\cite{OKA_T} is larger than the theoretical expectations quoted previously.

\section{The NA62 experiment at CERN}

\begin{figure}[h]
\centering
\includegraphics[width=.99\linewidth]{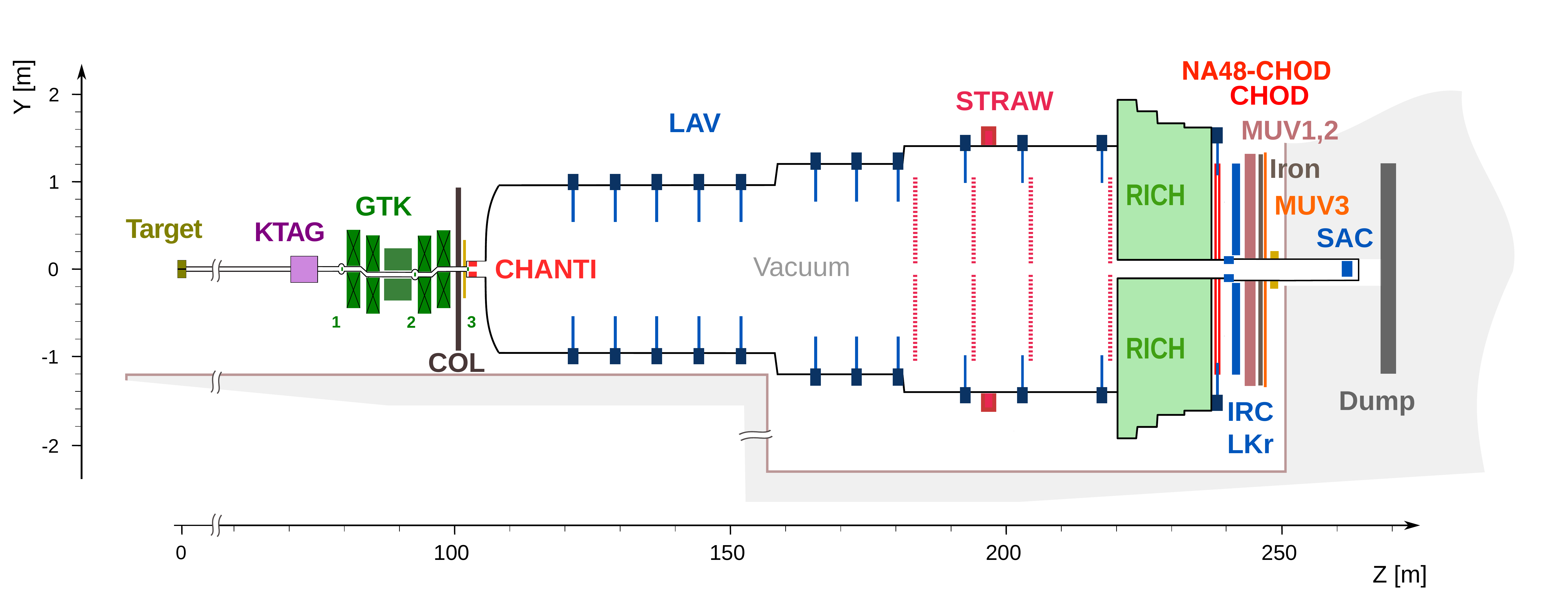}
\caption{
Schematic side view of the NA62 beam line and detector.
}
\label{fig:na62_scheme}
\end{figure}

The beam and detector of the NA62 experiment at the CERN SPS, designed to study the $K^+ \to \pi^+ \nu \Bar{\nu}$ decay~\cite{pinunu}, are described in~\cite{detector}.
A schematic view of the NA62 setup is presented in Figure~\ref{fig:na62_scheme}.

A 400~GeV/$c$ proton beam extracted from the SPS
impinges on the kaon production target in spills of three seconds effective duration.
The target position defines the origin of the NA62 reference system: the beam travels along the Z axis in the positive direction (downstream), the Y axis points vertically up, and the X axis is horizontal and directed to form a right-handed coordinate system.
Typical
intensities during data taking range from  $1.7$ to $1.9\times10^{12}$
protons per spill.  The resulting unseparated secondary hadron beam of positively
charged particles contains 70\% $\pi^+$, 23\% protons, 6\% $K^+$,
with a central beam momentum of 75~GeV/$c$ and 1\%~rms momentum bite.

Beam kaons are tagged with a time resolution of 70~ps by a differential Cherenkov counter (KTAG), with a 5~m long vessel containing nitrogen gas at 1.75~bar pressure, which acts as the radiator.
Beam particle positions, momenta and times (to better than 100 ps resolution) are measured  by a silicon pixel spectrometer consisting of three stations (GTK1,2,3) and four dipole magnets.
The typical beam particle rate at the third GTK station
is about 450~MHz.

The last station is immediately preceded by
a 1.2~m thick steel collimator (COL) with a 76~$\times$~40~mm$^2$
central aperture
and 1.7~$\times$~1.8~m$^2$ outer dimensions, to absorb hadrons from
upstream $K^+$ decays. A variable aperture collimator of about 0.15 $\times$~0.15~m$^2$ outer dimensions
was used up to early 2018.

The GTK3 station marks the beginning of a 117~m long vacuum tank.
In this analysis, a 60~m long fiducial volume (FV), in which 10\% of the kaons decay, is defined starting 7.6~m downstream of GTK3.
The beam has a  rectangular transverse profile of 52~$\times$~24 mm$^2$ and an angular spread of 0.11~mrad (rms) at the FV entrance.

The time, momentum and direction of charged particles produced by kaon decays
are measured by a magnetic spectrometer (STRAW).
The STRAW, comprised of two pairs of straw chambers on either side of a dipole magnet, measures momenta with a resolution
$\sigma_p / p = (0.30 \; \oplus \; 0.005 \cdot p)\%$,
where the momentum $p$ is expressed in GeV/$c$.
The ring-imaging Cherenkov counter (RICH), filled with neon at
atmospheric pressure, tags the decay particles with a time precision better than 100~ps and provides particle identification.
The CHOD, a matrix of tiles read out by Silicon photomultipliers, and the NA48-CHOD, composed of two orthogonal planes
of scintillating slabs,
are used for the trigger and time measurement, with a resolution of 200~ps.

Six stations of plastic scintillator bars (CHANTI) detect,
with 99\% efficiency and 1~ns time resolution, extra activity, including
inelastic interactions in  GTK3.
Twelve stations of ring-shaped electromagnetic calorimeters (LAV), made of lead-glass blocks, are located inside and downstream of the vacuum tank to achieve full acceptance for photons emitted by $K^+$ decays in the FV at polar angles between 10 and 50~mrad.
A 27 radiation-length thick, quasi-homogeneous liquid
krypton electromagnetic calorimeter (LKr) detects photons from $K^+$  decays
emitted at angles between 1 and 10~mrad. Its energy resolution is
$\sigma(E) / E = (4.8 / \sqrt{E} \; \oplus \; 11 / E \; \oplus \; 0.9)\%$, where the energy $E$ is expressed in GeV.  Its spatial and time
resolutions are 1~mm and between 0.5 and 1~ns, respectively, depending on the
amount and source of energy released.
The LKr also complements the RICH
particle identification.
Two hadronic iron/scintillator-strip
sampling calorimeters (MUV1,2) and an array of scintillator tiles located
behind 80 cm of iron (MUV3)  supplement the pion/muon separation system. 
The MUV3 provides a time resolution of 400~ps.  A lead/scintillator shashlik calorimeter
(IRC) located in front of the LKr and a detector based on the 
same principle (SAC) placed on the Z axis at
the downstream end of the apparatus, ensure the detection of photons down to
zero degrees in the forward direction.
The IRC and SAC detectors compose the Small Angle Veto system (SAV).
The LAV, LKr and SAV make the photon veto system almost hermetic for photons emitted by kaon decays in the FV:
the inefficiency of the detection of at least one photon from $K^+ \to \pi^+ \pi^0$, $\pi^0 \to \gamma \gamma$ decays is measured to be below $10^{-7}$, due to geometric acceptances and detector inefficiencies \cite{NA62:2020pwi}.

A two-level trigger system is used, with a hardware low level trigger, L0, and a software high level trigger, L1~\cite{readout, trigger}.
Auxiliary trigger lines are operated concurrently with the main trigger line that is dedicated to the $K^+ \to \pi^+ \nu \Bar{\nu}$ decay.
This analysis uses the \emph{non-$\mu$} and the \emph{control} trigger lines described below.
\begin{itemize}
    \item The \emph{non-$\mu$} line requires signals in the RICH and CHOD detectors at L0, compatible with the presence of at least one charged particle in the final state, and no signal in the MUV3 detector in time coincidence. A signal in the KTAG detector is required at L1, compatible with the presence of a charged kaon. In a fraction of the dataset, a track reconstructed by the STRAW is also required at L1. This trigger line is downscaled by a factor of 200.
    \item The \emph{control} line requires the presence of signal in the NA48-CHOD detector at L0, compatible with the presence of a charged particle in the final state. No requirements are applied at L1. This trigger line is downscaled by a factor of 400.
\end{itemize}

This analysis exploits the data collected by the NA62 experiment in 2017--2018.
Monte Carlo (MC) simulations of particle interactions with the detector and its response are performed using a software package based on the Geant4 toolkit~\cite{geant4}.

\section{Event selection}
\label{section:Selections}

Signal ($K_{e3\gamma}$) and normalization ($K_{e3}$) events share the same selection criteria, except for the requirement of an additional photon in the signal sample. This ensures a first-order cancellation of several systematic effects in the $R_j$ measurements.

\subsection{Common selection criteria}

A positively charged track is required to be reconstructed in the STRAW with momentum in the range 10--60~GeV/$c$.
Its extrapolated positions in the CHOD, NA48-CHOD, RICH and LKr front planes should be within the respective geometrical acceptances.
A spatial association of signals in these detectors is required, together with time association within 2~ns. Positron identification is achieved by applying conditions on the reconstructed RICH ring radius and on the ratio of the energy of the LKr cluster associated with the track to the measured track momentum.

The beam track is reconstructed in the GTK and identified as a kaon by an associated signal in KTAG within 2~ns.
The positron and kaon tracks are matched taking into account both space (closest distance of approach smaller than 10~mm) and time ($\pm 0.5$~ns) coincidence.
The mid-point of the segment at the closest distance of approach of the two tracks defines the kaon decay vertex, which is required to be within the FV.

The \emph{event time} ($T_{\rm event}$) is defined as the weighted average of the times measured by the KTAG, GTK, RICH and NA48-CHOD detectors, taking into account the resolutions of each detector,
and is required to be within 10~ns of the trigger time.

The two photons from $\pi^0 \to \gamma \gamma$ decay are identified by selecting two LKr clusters, not associated to any track, with energy above 4~GeV and within 3~ns of $T_{\rm event}$. The four-momentum of each photon is reconstructed using the energy and position of the cluster, assuming that the photon is produced at the kaon decay vertex. The di-photon mass is required to be compatible with the $\pi^0$ mass.
The $\pi^0$ four-momentum is defined as the sum of the four-momenta of the two photons.

The \emph{LKr time} ($T_{\rm LKr}$) is defined as the average of the times of the three LKr clusters associated with the positron and the photon pair forming the $\pi^0$.

Events with activity in the LAV and SAV within 15~ns of $T_{\rm event}$ are rejected to suppress background events coming mainly from \keqn decays. Events with signals in MUV3 within 15~ns of $T_{\rm event}$ are rejected to suppress background events with muons in the final state.

\subsection{Specific selection criteria}

The following exclusive criteria are applied to select signal and normalization candidates.

\paragraph{Normalization selection.}

The squared missing mass $m^2_{\rm miss}(K_{e3}) = (P_K - P_e - P_{\pi^0})^2$ must satisfy $|m^2_{\rm miss}(K_{e3})| < 11000$~MeV$^2/c^4$,
    where $P_K$, $P_e$ and $P_{\pi^0}$ are the reconstructed kaon, positron and $\pi^0$ four-momenta.
Events are rejected if a fourth LKr cluster is detected with energy above 2~GeV and within 15~ns of $T_{\rm LKr}$.

\paragraph{Signal selection.}

A fourth isolated cluster with energy above 4~GeV and time $T_\gamma$ within 3~ns of $T_{\rm LKr}$ must be present. It is identified as the radiative photon and attached to the kaon decay vertex.
Events are rejected if a fifth LKr cluster is detected with energy above 2~GeV and within 15~ns of $T_{\rm LKr}$. This condition further suppresses the \keqn background events.

The squared missing mass $m^2_{\rm miss}(K_{e3\gamma}) = (P_K - P_e - P_{\pi^0} -P_{\gamma})^2$  must satisfy $|m^2_{\rm miss}(K_{e3\gamma})| < 11000$~MeV$^2/c^4$ and  $m^2_{\rm miss}(K_{e3}) > 5000$~MeV$^2/c^4$, where $P_\gamma$ is the reconstructed four-momentum of the radiative photon.

The background from $K_{e3}$ events, with a bremsstrahlung photon emitted by positron interactions in the detector material, is suppressed by requiring a minimum distance between the radiative photon cluster
and the intersection of the positron line of flight at the vertex with the LKr plane.

Selection conditions on $E_{\gamma}$ and $\theta_{e \gamma}$
are applied according to the three kinematic region definitions, to obtain the signal sample $S_1$ and its subsets $S_2$ and $S_3$.

\section{Signal rate measurement}

The ratios $R_j$, defined in Eq.~(\ref{eq:br_def}), are measured as:
\begin{equation}
R_j = \frac{\mathcal{B}(K_{e3\gamma^j})}{\mathcal{B}(K_{e3})} = \frac{N^{\rm obs}_{Ke3\gamma^j}-N^{\rm bkg}_{Ke3\gamma^j}}{N^{\rm obs}_{Ke3}-N^{\rm bkg}_{Ke3}} \cdot \frac{A_{Ke3}}{A_{Ke3\gamma^j}} \cdot \frac{\epsilon^{\rm trig}_{Ke3}}{\epsilon^{\rm trig}_{Ke3\gamma^j}}.
    \label{eq:br_ke3g}
\end{equation}
Here, $N^{\rm obs}_{Ke3\gamma^j}$ and $N^{\rm bkg}_{Ke3\gamma^j}$ are the numbers of observed signal candidates and expected background events in the signal samples, and $A_{Ke3\gamma^j}$ and $\epsilon^{\rm trig}_{Ke3\gamma^j}$ are the related acceptances and the trigger efficiencies.
Similarly, $N^{\rm obs}_{Ke3}$ and $N^{\rm bkg}_{Ke3}$ are the numbers of observed normalization candidates and expected background events in the normalization sample, and $A_{Ke3}$ and $\epsilon^{\rm trig}_{Ke3}$ are the related acceptance and the trigger efficiency.

The signal selection acceptances are defined with respect to the corresponding kinematic regions, while the normalization selection acceptance is defined with respect to the full phase-space. They are evaluated using simulations. A ChPT $\mathcal{O}(p^6)$ description of the signal 
decay is used~\cite{Kubis:2006nh}. The normalization decay description includes only the IB component~\cite{Gatti:2005kw}. The 
resulting acceptances are reported in Table~\ref{tab:Nobs} together with the numbers of selected candidates.

The trigger conditions are replicated in the selections and are only related to the presence of charged particles.
The \emph{non-$\mu$} and \emph{control} trigger lines yield independent data samples. Each is used to evaluate the efficiency of the other.
The result is that the ratios $\epsilon^{\rm trig}_{Ke3} / \epsilon^{\rm trig}_{Ke3\gamma^j}$ are consistent with unity within 0.1\%.

\subsection{Background estimation}
\begin{figure}[h]
\centering
\includegraphics[width=.9\linewidth]{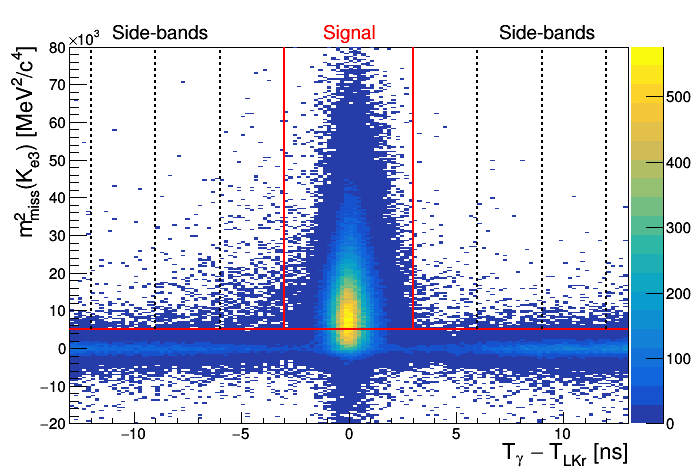}
\caption{
Reconstructed $m^2_{\rm miss}(K_{e3})$ of signal candidates as a function of the time difference $T_\gamma - T_{\rm LKr}$.
The $m^2_{\rm miss}(K_{e3})$ condition is shown (red horizontal line), together with the signal selection time window (red vertical lines) and the side-bands (black dashed vertical lines).
}
\label{fig:ke3g_acc}
\end{figure}

Table~\ref{tab:Nobs} summarises the sources of background considered.
The  only sizeable background in the normalization sample comes from \kdpi decays with the $\pi^+$ misidentified as $e^+$.
The main background in the signal sample comes from $K_{e3}$ and \kdpi decays (with $\pi^+$ misidentification in the latter case) with an extra cluster due to accidental activity in the LKr that mimics the radiative photon (accidental background).
The accidental background is measured from the data (Figure~\ref{fig:ke3g_acc}), using the side-bands $6< |T_{\gamma}-T_{\rm LKr}|<9 $~ns and assuming a flat $T_{\gamma}-T_{\rm LKr}$ distribution. This assumption is validated and the systematic uncertainty is evaluated using the validation side-bands $9< |T_{\gamma}-T_{\rm LKr}|<12 $~ns.
An additional contribution to the background in the signal sample stems from \keqn decays with a photon from the $\pi^0 \to \gamma \gamma$ decay not detected. Contributions from other potential background channels are found to be even lower than the $K^+ \to \pi^+ \pi^0 \pi^0$ estimation and therefore neglected.

\begin{figure}[h]
\centering
\includegraphics[width=.5\linewidth]{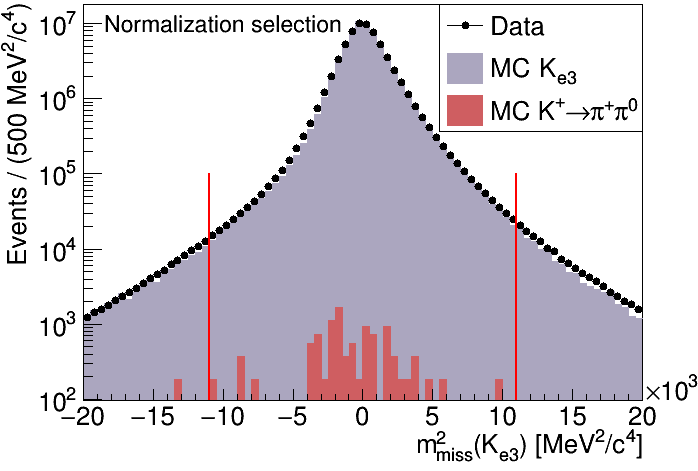}%
\includegraphics[width=.5\linewidth]{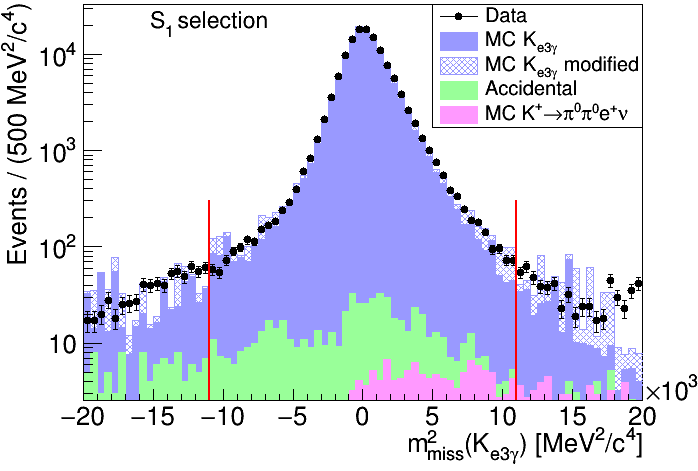} \\ 
\includegraphics[width=.5\linewidth]{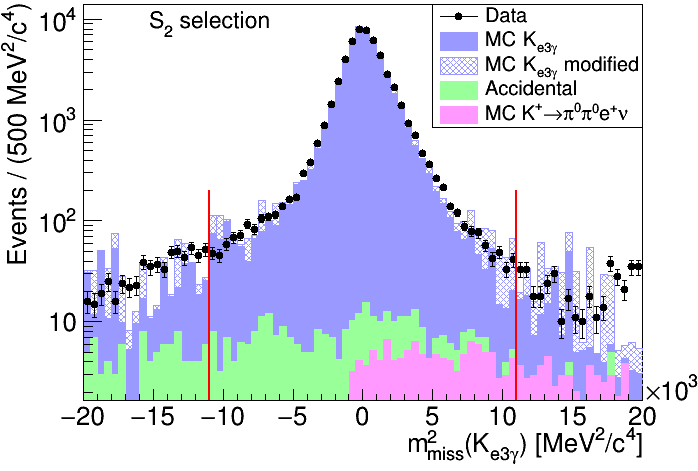}%
\includegraphics[width=.5\linewidth]{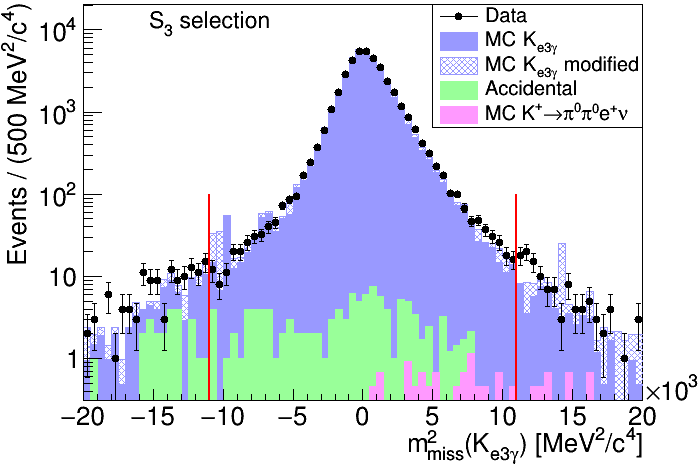}
\caption{
Top left: $m^2_{\rm miss}(K_{e3})$ distribution of the normalization sample for data (points) and expected signal and background (histograms).
Other panels: $m^2_{\rm miss}(K_{e3\gamma})$ distributions of the three signal samples for data (points) and expected signal and backgrounds (histograms); the
``$K_{e3\gamma}$ modified'' histogram represents the distribution after modification of the LKr response in the simulation (Section~\ref{sec:LKrSyst}).
The red lines correspond to the selection conditions.
}
\label{fig:mm2}
\end{figure}

All the background contributions, except for the accidental background, are estimated with simulations. The values of the branching fractions of the normalization and the background channels are taken from~\cite{pdg}.
The distributions of $m^2_{\rm miss}(K_{e3})$ for the selected normalization events, and of $m^2_{\rm miss}(K_{e3\gamma})$ for the selected signal events, are shown in Figure~\ref{fig:mm2} for the data, estimated background and simulated signal and normalization decays.

\begin{table}[h]
\caption{
Numbers of selected candidates, acceptances, background estimates
and fractional background
for the normalization and the three signal samples. The fractional background is the ratio of the total background to the number of selected candidates. The accidental background does not apply to the normalization sample.
The \kdpi background in the signal samples is included in the accidental background.
}
 \scriptsize
\begin{center}
\begin{tabular}{|l||c|c|c|c|}
\hline
 & Normalization & $S_1$ & $S_2$ & $S_3$ \\ \hline \hline
 Selected candidates & $6.6420\times 10^7$ & $1.2966 \times 10^5$ & $0.5359 \times 10^5$ & $0.3909 \times 10^5$ \\ \hline
Acceptance & $(3.842 \pm 0.002)\%$ & $(0.444 \pm 0.001)\%$ & $(0.514 \pm 0.002)\%$ & $(0.432 \pm 0.002)\%$ \\ \hline
Accidental background & --- & $ (4.9 \pm 0.2 \pm 1.3) \times 10^2 $ & $ (2.3 \pm 0.2 \pm 0.3) \times 10^2 $ & $ (1.1 \pm 0.1 \pm 0.5) \times 10^2 $  \\
$K^+ \to \pi^0 \pi^0 e^+ \nu$ & $<10^2$ & $ (1.1 \pm 1.1) \times 10^2 $ & $ (1.1 \pm 1.1) \times 10^2 $ & $ (0.1 \pm 0.1) \times 10^2 $  \\
$K^+ \to \pi^+ \pi^0 \pi^0 $ & $<10^2$ & $ <20 $ & $ <20 $ & $ <20 $  \\
$K^+ \to \pi^+ \pi^0 $ & $(1.0 \pm 1.0) \times 10^4$ & --- & --- & --- \\ \hline
Total background & $ (1.0 \pm 1.0) \times 10^4 $ & $ (6.0 \pm 1.8) \times 10^2 $ & $ (3.4 \pm 1.2) \times 10^2 $ & $ (1.2 \pm 0.6) \times 10^2 $  \\ \hline
Fractional background & $ 1.6 \times 10^{-4} $ & $ 0.46 \times 10^{-2} $ & $ 0.64 \times 10^{-2} $ & $ 0.29 \times 10^{-2}$  \\ \hline
\end{tabular}
\end{center}
\label{tab:Nobs}
\end{table}

\subsection{LKr response correction}
\label{sec:LKrSyst}
The signal selection acceptance is sensitive to the precision of the low-energy photon measurement in the LKr due to the steeply falling radiative photon energy spectrum. The minimum radiative photon energy considered in the analysis is 4~GeV, while the standard LKr fine calibration procedures exploit $\pi^0\to\gamma\gamma$ and $\pi^0\to\gamma e^+e^-$ decays, and provide optimal precision in the 10--30~GeV energy range. A dedicated procedure is therefore employed to modify the LKr response in the simulated samples and estimate the corresponding systematic uncertainty. The procedure is based on improving the agreement between data and simulated distributions in the most relevant variables, $m^2_{\rm miss}(K_{e3})$ and $m^2_{\rm miss}(K_{e3\gamma})$.
The LKr response modification includes a constant energy scale factor, an energy-dependent factor, and an additional stochastic smearing of the measured energy.
The scale-related uncertainty is the largest contribution in $R_1$ and $R_3$; the stochastic smearing uncertainty is of similar magnitude and is the major contribution in $R_2$.

The correction factors $f_j^{\rm LKr}$ are defined as the ratios of the corrected $A_{Ke3}/A_{Ke3\gamma^j}$ values to the original values. The systematic uncertainties in $f_j^{\rm LKr}$ are evaluated as quadratic sums of three terms: the maximum deviations of $f_j^{\rm LKr}$ from their central values obtained by varying the modification parameters within their uncertainties; half of the size of the overall correction; and half of the maximum variation of $f_j^{\rm LKr}$ obtained using alternative LKr fine calibration procedures based on different methods of $\pi^0$ mass reconstruction. The correction factors and their uncertainties are listed in Table~\ref{fPVcorr}.
The acceptance ratios $A_{Ke3}/A_{Ke3\gamma^j}$ (Table~\ref{tab:Nobs}) must be multiplied by $f_j^{\rm LKr}$.

\subsection{Photon veto correction}

The signal and normalization MC generators, used to evaluate acceptance, simulate exactly one radiative photon per event.
Both the normalization and signal selections forbid the presence of in-time extra clusters in the LKr with 
energy above $2$~GeV, as well as in-time signals in the LAV and SAV.
The radiative photon contributes to the photon veto simulation in the normalization case, but not in the  signal case where it is included in the signal event reconstruction. Additional radiative photons are not generated in the simulated samples, leading to a systematic underestimation of the acceptance ratios $A_{Ke3}/A_{Ke3\gamma^j}$.
However, special signal and normalization MC samples with multiple radiative photons are generated using the PHOTOS program~\cite{Barberio:1993qi}, providing a less accurate description of the radiative effects than the standard MC samples and therefore used only to evaluate the photon veto effects.
The resulting correction factors $f_j^{\rm PV}$ are reported in Table~\ref{fPVcorr}.
The acceptance ratios $A_{Ke3}/A_{Ke3\gamma^j}$ (Table~\ref{tab:Nobs}) must be multiplied by $f_j^{\rm PV}$.

The uncertainties in $f_j^{\rm PV}$ include MC statistical uncertainties, variations 
of the simulated LAV and SAV efficiencies when replaced by the measured energy-dependent 
values (evaluated with data using photons from $\pi^0 \to \gamma \gamma$ decays)~\cite{NA62:2020pwi}, and the variation of the LKr veto efficiency with the LKr response tuning.
The deviation of $f_j^{\rm PV}$ from unity is dominated by the LKr contribution, while the uncertainties in $f_j^{\rm PV}$ are dominated by the LAV contributions.

\begin{table}[h]
\caption{
Correction factors to the acceptance ratios $A_{Ke3}/A_{Ke3\gamma^j}$
from LKr response modelling ($f_j^{\rm LKr}$) and photon veto  ($f_j^{\rm PV}$), for the three signal samples.
}
\label{fPVcorr}
\begin{center}
\begin{tabular}{|l||c|c|c|} \hline
          & $S_1$ &  $S_2$ &  $S_3$ \\ \hline \hline
$f_j^{\rm LKr}$  & {\ensuremath{0.9967 \pm 0.0038}\xspace}  & {\ensuremath{0.9941 \pm 0.0047}\xspace}  & {\ensuremath{0.9996 \pm 0.0039}\xspace} \\  \hline
$f_j^{\rm PV}$   & {\ensuremath{1.0234 \pm 0.0028}\xspace}  & {\ensuremath{1.0243 \pm 0.0036}\xspace}  & {\ensuremath{1.0216 \pm 0.0031}\xspace} \\ \hline
\end{tabular}
\end{center}
\end{table}

\subsection{Theoretical model uncertainty}
The theoretical model used in the MC simulation of the $K_{e3\gamma}$ decay, based on ChPT $\mathcal{O}(p^6)$, results in a 30\% relative uncertainty in the contribution to the decay width arising from the SD component and its interference with the IB component~\cite{Kubis:2006nh}.
The acceptances evaluated with the signal MC sample are compared with those evaluated with the $K_{e3}$ MC sample generated including only the IB component of the radiative effects~\cite{Gatti:2005kw}, and 30\% of their relative difference is considered as a relative systematic uncertainty in $A_{Ke3\gamma^j}$.

\subsection{Results}
The measured $R_j$ values are:
\begin{equation*}
\begin{split}
    & R_1 \times 10^2 = 1.715 \pm 0.005_{\text{stat}} \pm 0.010_{\text{syst}} = 1.715 \pm 0.011, \\
    & R_2 \times 10^2 = 0.609 \pm 0.003_{\text{stat}} \pm 0.006_{\text{syst}} = 0.609 \pm 0.006, \\
    & R_3 \times 10^2 = 0.533 \pm 0.003_{\text{stat}} \pm 0.004_{\text{syst}} = 0.533 \pm 0.004.
\end{split}
\label{eq:Rj}
\end{equation*}
The error budgets are given in Table~\ref{tab:errorBudget}.
The statistical uncertainties quoted arise from the numbers of observed candidates in the data, while all other contributions are considered as systematic uncertainties.
The stability of the results is checked by splitting the data sample into subsamples, and by varying the selection conditions, with no evidence for residual systematic effects.

\begin{table}[h]
\caption{Relative uncertainties in the $R_j$ measurements.}
\begin{center}
\begin{tabular}{|l||c|c|c|}
\hline
 & $\delta R_1/R_1$ & $\delta R_2/R_2$ & $\delta R_3/R_3$ \\ \hline \hline
Statistical & 0.3\% & 0.4\% & 0.5\% \\ \hline
Limited MC sample size & 0.2\% & 0.4\% & 0.4\% \\
Background estimation & 0.1\% & 0.2\% & 0.1\% \\
LKr response modelling & 0.4\% & 0.5\% & 0.4\% \\ 
Photon veto correction & 0.3\%  & 0.4\% & 0.3\% \\ 
Theoretical model & 0.1\% & 0.5\% & 0.1\% \\ \hline
Total systematic & 0.6\% & 0.9\% & 0.7\% \\ \hline
Total & 0.7\% & 1.0\% & 0.8\% \\ \hline
\end{tabular}
\end{center}
\label{tab:errorBudget}
\end{table}

The measured values are $5\%$ smaller than the ChPT $\mathcal{O}(p^6)$ calculations reported in Table~\ref{tab:ke3g_stateofart}, with a disagreement at the level of three standard deviations.
The smaller and less precise value of $R_2$ given in~\cite{Khriplovich:2010rz} is also different from the measurement at the level of three standard deviations.

\section{T-asymmetry measurement}
\begin{figure}[h]
\centering
\includegraphics[width=.5\linewidth]{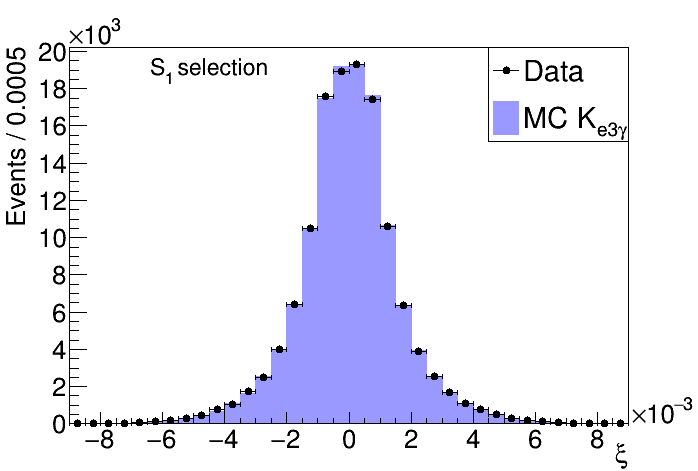}%
\includegraphics[width=.5\linewidth]{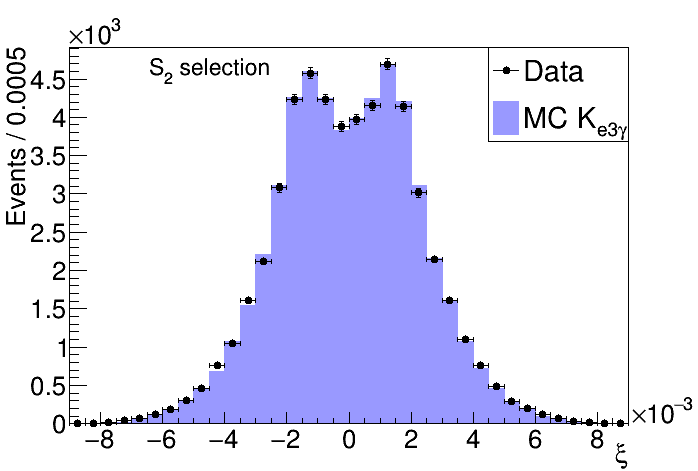} \\
\includegraphics[width=.5\linewidth]{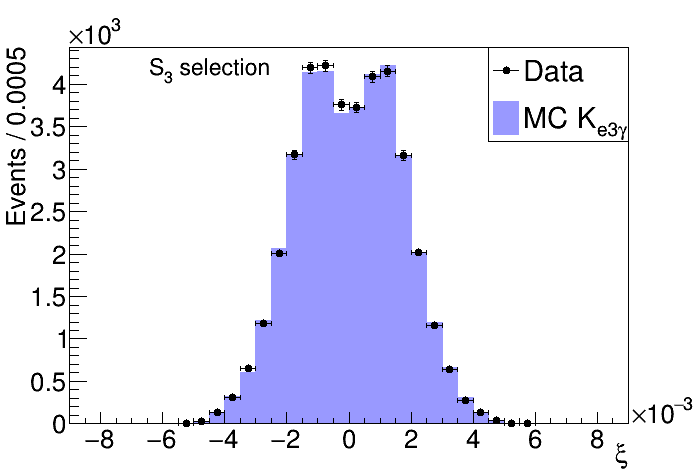}%
\caption{
Distributions of the reconstructed $\xi$ observable for the three signal samples for data (points) and simulation (histograms).
}
\label{fig:xi}
\end{figure}

The T-asymmetry $A_\xi$ is measured using the $K_{e3\gamma}$ samples selected for the $R_j$ measurements.
The distributions of the $\xi$ observable are shown in Figure~\ref{fig:xi}, for data and simulation.

For each selected signal sample, a raw asymmetry measurement $A_\xi^{\rm Data}$ is obtained using Eq.~(\ref{eq:def_xi}).
An offset $A_\xi^{\rm MC}$, possibly introduced by the detector and the selection, is measured by comparing the reconstructed and generated asymmetry values in the $K_{e3\gamma}$ simulated sample.
The generated asymmetry value is checked to be zero within a $\mathcal{O}(10^{-4})$ precision in the three $K_{e3\gamma}$ kinematic regions considered, in agreement with the ChPT $\mathcal{O}(p^6)$ calculations~\cite{Muller:2006gu}.
The final measurement is obtained as $A_\xi^{\rm NA62} = A_\xi^{\rm Data}-A_\xi^{\rm MC}$. The uncertainty in $A_\xi^{\rm MC}$, due to the limited statistics of the $K_{e3\gamma}$ simulated sample, is propagated as a systematic uncertainty.
No statistically significant asymmetry is observed, as reported in Table~\ref{tab:Tasymm_res}.

\begin{table}[h]
\caption{
Results of the $A_\xi$ measurements for the three signal samples. The measurements by the OKA experiment~\cite{OKA_T}, $A_\xi^{\rm OKA}$, are also reported for comparison. The sign of the OKA results is changed for consistency with Eq.~(\ref{eq:def_xi}).
}
\begin{center}
\begin{tabular}{|ll||c|c|c|}
\hline
 & & $S_1$ & $S_2$ & $S_3$  \\ \hline \hline
$A_\xi^{\rm Data}$&$ \times 10^3$ & $1.9 \pm 2.8_{\text{stat}}$ & $1.3 \pm 4.3_{\text{stat}}$ & $-6.2 \pm 5.1_{\text{stat}}$ \\
$A_\xi^{\rm MC}$&$ \times 10^3$ & $3.1 \pm 1.9_{\text{syst}}$ & $4.8 \pm 3.0_{\text{syst}}$ & \phantom{0} $3.0 \pm 3.5_{\text{syst}}$ \\ \hline
$A_\xi^{\rm NA62}$&$ \times 10^3$ & $-1.2 \pm 2.8_{\text{stat}} \pm 1.9_{\text{syst}}$ & $-3.4 \pm 4.3_{\text{stat}} \pm 3.0_{\text{syst}}$ & $-9.1 \pm 5.1_{\text{stat}} \pm 3.5_{\text{syst}}$ \\ \hline
$A_\xi^{\rm OKA}$&$ \times 10^3$ & $-0.1 \pm 3.9_{\text{stat}} \pm 1.7_{\text{syst}}$ & $-4.4 \pm 7.9_{\text{stat}} \pm 1.9_{\text{syst}}$ & \phantom{0} $7.0 \pm 8.1_{\text{stat}} \pm 1.5_{\text{syst}}$ \\ \hline
\end{tabular}
\end{center}
\label{tab:Tasymm_res}
\end{table}

\section{Summary}

Measurements of the ratio of the \ketg to \ket branching fractions, together with measurements of the T-violating asymmetry in the \ketg decay, are performed in three kinematic regions using data collected by the NA62 experiment at CERN in 2017--2018.

The measured ratios,
$R_1 = (1.715 \pm 0.011)\times 10^{-2}$,
$R_2 = (0.609 \pm 0.006)\times 10^{-2}$,
$R_3 = (0.533 \pm 0.004)\times 10^{-2}$,
are at least a factor of two more precise than previous measurements. The relative uncertainties do not exceed 1\%, matching the precision of the most precise theoretical calculations.

The T-asymmetry measurements performed at an improved precision are compatible with no asymmetry in the three kinematic regions considered. Their uncertainties remain larger than the theoretical expectations.

\newpage
\clearpage
\section*{Acknowledgements}
It is a pleasure to express our appreciation to the staff of the CERN laboratory and the technical
staff of the participating laboratories and universities for their efforts in the operation of the
experiment and data processing.

The cost of the experiment and its auxiliary systems was supported by the funding agencies of 
the Collaboration Institutes. We are particularly indebted to: 
F.R.S.-FNRS (Fonds de la Recherche Scientifique - FNRS), under Grants No. 4.4512.10, 1.B.258.20, Belgium;
CECI (Consortium des Equipements de Calcul Intensif), funded by the Fonds de la Recherche Scientifique de Belgique (F.R.S.-FNRS) under Grant No. 2.5020.11 and by the Walloon Region, Belgium;
NSERC (Natural Sciences and Engineering Research Council), funding SAPPJ-2018-0017,  Canada;
MEYS (Ministry of Education, Youth and Sports) funding LM 2018104, Czech Republic;
BMBF (Bundesministerium f\"{u}r Bildung und Forschung) contracts 05H12UM5, 05H15UMCNA and 05H18UMCNA, Germany;
INFN  (Istituto Nazionale di Fisica Nucleare),  Italy;
MIUR (Ministero dell'Istruzione, dell'Universit\`a e della Ricerca),  Italy;
CONACyT  (Consejo Nacional de Ciencia y Tecnolog\'{i}a),  Mexico;
IFA (Institute of Atomic Physics) Romanian 
CERN-RO No. 1/16.03.2016 
and Nucleus Programme PN 19 06 01 04,  Romania;
MESRS  (Ministry of Education, Science, Research and Sport), Slovakia; 
CERN (European Organization for Nuclear Research), Switzerland; 
STFC (Science and Technology Facilities Council), United Kingdom;
NSF (National Science Foundation) Award Numbers 1506088 and 1806430,  U.S.A.;
ERC (European Research Council)  ``UniversaLepto'' advanced grant 268062, ``KaonLepton'' starting grant 336581, Europe.

Individuals have received support from:
Charles University Research Center (UNCE/SCI/013), Czech Republic;
Ministero dell'Istruzione, dell'Universit\`a e della Ricerca (MIUR  ``Futuro in ricerca 2012''  grant RBFR12JF2Z, Project GAP), Italy;
the Royal Society  (grants UF100308, UF0758946), United Kingdom;
STFC (Rutherford fellowships ST/J00412X/1, ST/M005798/1), United Kingdom;
ERC (grants 268062,  336581 and  starting grant 802836 ``AxScale'');
EU Horizon 2020 (Marie Sk\l{}odowska-Curie grants 701386, 754496, 842407, 893101, 101023808).


\newpage
\clearpage

\newcommand{\orcimg}{\raisebox{-0.3\height}{\includegraphics[height=\fontcharht\font`A]{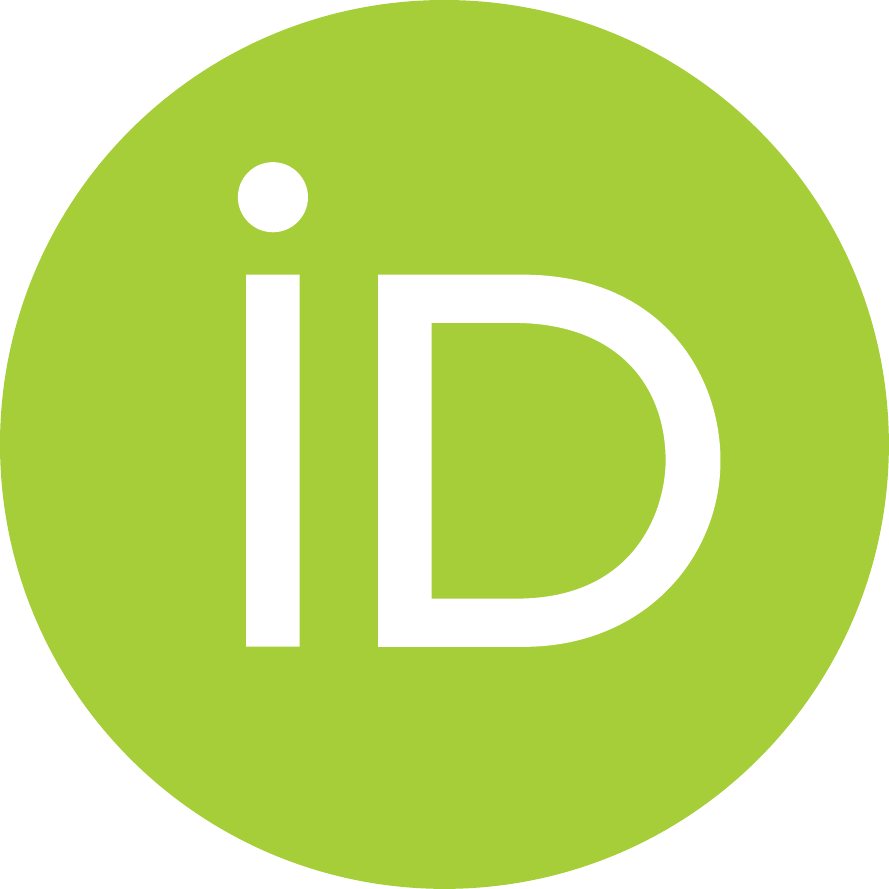}}}
\newcommand{\orcid}[1]{\href{https://orcid.org/#1}{\orcimg}}

\centerline{\bf The NA62 Collaboration} 
\vspace{0.5cm}
%
%

\begin{raggedright}
\noindent
{\bf Universit\'e Catholique de Louvain, Louvain-La-Neuve, Belgium}\\
 E.~Cortina Gil\orcid{0000-0001-9627-699X},
 A.~Kleimenova$\,${\footnotemark[1]}\orcid{0000-0002-9129-4985},
 E.~Minucci$\,${\footnotemark[2]}\orcid{0000-0002-3972-6824},
 S.~Padolski\orcid{0000-0002-6795-7670},
 P.~Petrov, 
 A.~Shaikhiev$\,${\footnotemark[3]}\orcid{0000-0003-2921-8743},
 R.~Volpe$\,${\footnotemark[4]}\orcid{0000-0003-1782-2978}
\vspace{0.5cm}

{\bf TRIUMF, Vancouver, British Columbia, Canada}\\
 T.~Numao\orcid{0000-0001-5232-6190},
 Y.~Petrov\orcid{0000-0003-2643-8740},
 B.~Velghe\orcid{0000-0002-0797-8381},
 V. W. S.~Wong\orcid{0000-0001-5975-8164}
\vspace{0.5cm}

{\bf University of British Columbia, Vancouver, British Columbia, Canada}\\
 D.~Bryman$\,${\footnotemark[5]}\orcid{0000-0002-9691-0775},
 J.~Fu
\vspace{0.5cm}

{\bf Charles University, Prague, Czech Republic}\\
 Z.~Hives\orcid{0000-0002-5025-993X},
 T.~Husek\orcid{0000-0002-7208-9150},
 J.~Jerhot$\,${\footnotemark[6]}\orcid{0000-0002-3236-1471},
 K.~Kampf\orcid{0000-0003-1096-667X},
 M.~Zamkovsky$\,${\footnotemark[7]}\orcid{0000-0002-5067-4789}
\vspace{0.5cm}

{\bf Aix Marseille University, CNRS/IN2P3, CPPM, Marseille, France}\\
 B.~De Martino\orcid{0000-0003-2028-9326},
 M.~Perrin-Terrin\orcid{0000-0002-3568-1956}
\vspace{0.5cm}

{\bf Institut f\"ur Physik and PRISMA Cluster of Excellence, Universit\"at Mainz, Mainz, Germany}\\
 A.T.~Akmete\orcid{0000-0002-5580-5477},
 R.~Aliberti$\,${\footnotemark[8]}\orcid{0000-0003-3500-4012},
 G.~Khoriauli$\,${\footnotemark[9]}\orcid{0000-0002-6353-8452},
 J.~Kunze,
 D.~Lomidze$\,${\footnotemark[10]}\orcid{0000-0003-3936-6942}, 
 L.~Peruzzo\orcid{0000-0002-4752-6160},
 M.~Vormstein,
 R.~Wanke\orcid{0000-0002-3636-360X}
\vspace{0.5cm}

{\bf Dipartimento di Fisica e Scienze della Terra dell'Universit\`a e INFN, Sezione di Ferrara, Ferrara, Italy}\\
 P.~Dalpiaz,
 M.~Fiorini\orcid{0000-0001-6559-2084},
 A.~Mazzolari\orcid{0000-0003-0804-6778},
 I.~Neri\orcid{0000-0002-9669-1058},
 A.~Norton$\,${\footnotemark[11]}\orcid{0000-0001-5959-5879}, 
 F.~Petrucci\orcid{0000-0002-7220-6919},
 M.~Soldani\orcid{0000-0003-4902-943X},
 H.~Wahl$\,${\footnotemark[12]}\orcid{0000-0003-0354-2465}
\vspace{0.5cm}

{\bf INFN, Sezione di Ferrara, Ferrara, Italy}\\
 L.~Bandiera\orcid{0000-0002-5537-9674},
 A.~Cotta Ramusino\orcid{0000-0003-1727-2478},
 A.~Gianoli\orcid{0000-0002-2456-8667},
 M.~Romagnoni\orcid{0000-0002-2775-6903},
 A.~Sytov\orcid{0000-0001-8789-2440}
\vspace{0.5cm}

{\bf Dipartimento di Fisica e Astronomia dell'Universit\`a e INFN, Sezione di Firenze, Sesto Fiorentino, Italy}\\
 E.~Iacopini\orcid{0000-0002-5605-2497},
 G.~Latino\orcid{0000-0002-4098-3502},
 M.~Lenti\orcid{0000-0002-2765-3955},
 P.~Lo Chiatto\orcid{0000-0002-4177-557X},
 I.~Panichi\orcid{0000-0001-7749-7914},
 A.~Parenti\orcid{0000-0002-6132-5680}
\vspace{0.5cm}

{\bf INFN, Sezione di Firenze, Sesto Fiorentino, Italy}\\
 A.~Bizzeti$\,${\footnotemark[13]}\orcid{0000-0001-5729-5530},
 F.~Bucci\orcid{0000-0003-1726-3838}
\vspace{0.5cm}

{\bf Laboratori Nazionali di Frascati, Frascati, Italy}\\
 A.~Antonelli\orcid{0000-0001-7671-7890},
 G.~Georgiev$\,${\footnotemark[14]}\orcid{0000-0001-6884-3942},
 V.~Kozhuharov$\,${\footnotemark[14]}\orcid{0000-0002-0669-7799},
 G.~Lanfranchi\orcid{0000-0002-9467-8001},
 S.~Martellotti\orcid{0000-0002-4363-7816}, 
 M.~Moulson\orcid{0000-0002-3951-4389},
 T.~Spadaro\orcid{0000-0002-7101-2389},
 G.~Tinti\orcid{0000-0003-1364-844X}
\vspace{0.5cm}

{\bf Dipartimento di Fisica ``Ettore Pancini'' e INFN, Sezione di Napoli, Napoli, Italy}\\
 F.~Ambrosino\orcid{0000-0001-5577-1820},
 T.~Capussela,
 M.~Corvino\orcid{0000-0002-2401-412X},
 M.~D'Errico\orcid{0000-0001-5326-1106},
 D.~Di Filippo\orcid{0000-0003-1567-6786}, 
 R.~Fiorenza$\,${\footnotemark[15]}\orcid{0000-0003-4965-7073},
 R.~Giordano\orcid{0000-0002-5496-7247},
 P.~Massarotti\orcid{0000-0002-9335-9690},
 M.~Mirra\orcid{0000-0002-1190-2961},
 M.~Napolitano\orcid{0000-0003-1074-9552}, 
 I.~Rosa\orcid{0009-0002-7564-182},
 G.~Saracino\orcid{0000-0002-0714-5777}
\vspace{0.5cm}

{\bf Dipartimento di Fisica e Geologia dell'Universit\`a e INFN, Sezione di Perugia, Perugia, Italy}\\
 G.~Anzivino\orcid{0000-0002-5967-0952},
 F.~Brizioli$\,$\renewcommand{\thefootnote}{\fnsymbol{footnote}}\footnotemark[1]\renewcommand{\thefootnote}{\arabic{footnote}}$^,$$\,${\footnotemark[7]}\orcid{0000-0002-2047-441X},
 E.~Imbergamo,
 R.~Lollini\orcid{0000-0003-3898-7464},
 R.~Piandani$\,${\footnotemark[16]}\orcid{0000-0003-2226-8924},
 C.~Santoni\orcid{0000-0001-7023-7116}
\vspace{0.5cm}

{\bf INFN, Sezione di Perugia, Perugia, Italy}\\
 M.~Barbanera\orcid{0000-0002-3616-3341},
 P.~Cenci\orcid{0000-0001-6149-2676},
 B.~Checcucci\orcid{0000-0002-6464-1099},
 P.~Lubrano\orcid{0000-0003-0221-4806},
 M.~Lupi$\,${\footnotemark[17]}\orcid{0000-0001-9770-6197}, 
 M.~Pepe\orcid{0000-0001-5624-4010},
 M.~Piccini\orcid{0000-0001-8659-4409}
\vspace{0.5cm}

{\bf Dipartimento di Fisica dell'Universit\`a e INFN, Sezione di Pisa, Pisa, Italy}\\
 F.~Costantini\orcid{0000-0002-2974-0067},
 L.~Di Lella$\,${\footnotemark[12]}\orcid{0000-0003-3697-1098},
 N.~Doble$\,${\footnotemark[12]}\orcid{0000-0002-0174-5608},
 M.~Giorgi\orcid{0000-0001-9571-6260},
 S.~Giudici\orcid{0000-0003-3423-7981}, 
 G.~Lamanna\orcid{0000-0001-7452-8498},
 E.~Lari\orcid{0000-0003-3303-0524},
 E.~Pedreschi\orcid{0000-0001-7631-3933},
 M.~Sozzi\orcid{0000-0002-2923-1465}
\vspace{0.5cm}

{\bf INFN, Sezione di Pisa, Pisa, Italy}\\
 C.~Cerri,
 R.~Fantechi\orcid{0000-0002-6243-5726},
 L.~Pontisso$\,${\footnotemark[18]}\orcid{0000-0001-7137-5254},
 F.~Spinella\orcid{0000-0002-9607-7920}
\vspace{0.5cm}

{\bf Scuola Normale Superiore e INFN, Sezione di Pisa, Pisa, Italy}\\
 I.~Mannelli\orcid{0000-0003-0445-7422}
\vspace{0.5cm}

{\bf Dipartimento di Fisica, Sapienza Universit\`a di Roma e INFN, Sezione di Roma I, Roma, Italy}\\
 G.~D'Agostini\orcid{0000-0002-6245-875X},
 M.~Raggi\orcid{0000-0002-7448-9481}
\vspace{0.5cm}

{\bf INFN, Sezione di Roma I, Roma, Italy}\\
 A.~Biagioni\orcid{0000-0001-5820-1209},
 P.~Cretaro\orcid{0000-0002-2229-149X},
 O.~Frezza\orcid{0000-0001-8277-1877},
 E.~Leonardi\orcid{0000-0001-8728-7582},
 A.~Lonardo\orcid{0000-0002-5909-6508}, 
 M.~Turisini\orcid{0000-0002-5422-1891},
 P.~Valente\orcid{0000-0002-5413-0068},
 P.~Vicini\orcid{0000-0002-4379-4563}
\vspace{0.5cm}

{\bf INFN, Sezione di Roma Tor Vergata, Roma, Italy}\\
 R.~Ammendola\orcid{0000-0003-4501-3289},
 V.~Bonaiuto$\,${\footnotemark[19]}\orcid{0000-0002-2328-4793},
 A.~Fucci,
 A.~Salamon\orcid{0000-0002-8438-8983},
 F.~Sargeni$\,${\footnotemark[20]}\orcid{0000-0002-0131-236X}
\vspace{0.5cm}

{\bf Dipartimento di Fisica dell'Universit\`a e INFN, Sezione di Torino, Torino, Italy}\\
 R.~Arcidiacono$\,${\footnotemark[21]}\orcid{0000-0001-5904-142X},
 B.~Bloch-Devaux\orcid{0000-0002-2463-1232},
 M.~Boretto$\,${\footnotemark[7]}\orcid{0000-0001-5012-4480},
 E.~Menichetti\orcid{0000-0001-7143-8200},
 E.~Migliore\orcid{0000-0002-2271-5192},
 D.~Soldi\orcid{0000-0001-9059-4831}
\vspace{0.5cm}

{\bf INFN, Sezione di Torino, Torino, Italy}\\
 C.~Biino\orcid{0000-0002-1397-7246},
 A.~Filippi\orcid{0000-0003-4715-8748},
 F.~Marchetto\orcid{0000-0002-5623-8494}
\vspace{0.5cm}

{\bf Instituto de F\'isica, Universidad Aut\'onoma de San Luis Potos\'i, San Luis Potos\'i, Mexico}\\
 A.~Briano Olvera\orcid{0000-0001-6121-3905},
 J.~Engelfried\orcid{0000-0001-5478-0602},
 N.~Estrada-Tristan$\,${\footnotemark[22]}\orcid{0000-0003-2977-9380},
 M. A.~Reyes Santos$\,${\footnotemark[22]}\orcid{0000-0003-1347-2579}
\vspace{0.5cm}

{\bf Horia Hulubei National Institute for R\&D in Physics and Nuclear Engineering, Bucharest-Magurele, Romania}\\
 P.~Boboc\orcid{0000-0001-5532-4887},
 A. M.~Bragadireanu,
 S. A.~Ghinescu\orcid{0000-0003-3716-9857},
 O. E.~Hutanu
\vspace{0.5cm}

{\bf Faculty of Mathematics, Physics and Informatics, Comenius University, Bratislava, Slovakia}\\
 L.~Bician$\,${\footnotemark[23]}\orcid{0000-0001-9318-0116},
 T.~Blazek\orcid{0000-0002-2645-0283},
 V.~Cerny\orcid{0000-0003-1998-3441},
 Z.~Kucerova$\,${\footnotemark[7]}\orcid{0000-0001-8906-3902}
\vspace{0.5cm}

{\bf CERN, European Organization for Nuclear Research, Geneva, Switzerland}\\
 J.~Bernhard\orcid{0000-0001-9256-971X},
 A.~Ceccucci\orcid{0000-0002-9506-866X},
 M.~Ceoletta\orcid{0000-0002-2532-0217},
 H.~Danielsson\orcid{0000-0002-1016-5576},
 N.~De Simone$\,${\footnotemark[24]}, 
 F.~Duval,
 B.~D\"obrich$\,${\footnotemark[25]}\orcid{0000-0002-6008-8601},
 L.~Federici\orcid{0000-0002-3401-9522},
 E.~Gamberini\orcid{0000-0002-6040-4985},
 L.~Gatignon$\,${\footnotemark[3]}\orcid{0000-0001-6439-2945}, 
 R.~Guida,
 F.~Hahn$\,$\renewcommand{\thefootnote}{\fnsymbol{footnote}}\footnotemark[2]\renewcommand{\thefootnote}{\arabic{footnote}},
 E.~B.~Holzer\orcid{0000-0003-2622-6844},
 B.~Jenninger,
 M.~Koval$\,${\footnotemark[23]}\orcid{0000-0002-6027-317X}, 
 P.~Laycock$\,${\footnotemark[26]}\orcid{0000-0002-8572-5339},
 G.~Lehmann Miotto\orcid{0000-0001-9045-7853},
 P.~Lichard\orcid{0000-0003-2223-9373},
 A.~Mapelli\orcid{0000-0002-4128-1019},
 R.~Marchevski$\,${\footnotemark[1]}\orcid{0000-0003-3410-0918}, 
 K.~Massri\orcid{0000-0001-7533-6295},
 M.~Noy,
 V.~Palladino\orcid{0000-0002-9786-9620},
 J.~Pinzino$\,${\footnotemark[27]}\orcid{0000-0002-7418-0636},
 V.~Ryjov, 
 S.~Schuchmann\orcid{0000-0002-8088-4226},
 S.~Venditti
\vspace{0.5cm}

{\bf School of Physics and Astronomy, University of Birmingham, Birmingham, United Kingdom}\\
 T.~Bache\orcid{0000-0003-4520-830X},
 M. B.~Brunetti$\,${\footnotemark[28]}\orcid{0000-0003-1639-3577},
 V.~Duk$\,${\footnotemark[4]}\orcid{0000-0001-6440-0087},
 V.~Fascianelli$\,${\footnotemark[29]},
 J. R.~Fry\orcid{0000-0002-3680-361X}, 
 F.~Gonnella\orcid{0000-0003-0885-1654},
 E.~Goudzovski\orcid{0000-0001-9398-4237},
 J.~Henshaw\orcid{0000-0001-7059-421X},
 L.~Iacobuzio,
 C.~Kenworthy\orcid{0009-0002-8815-0048}, 
 C.~Lazzeroni\orcid{0000-0003-4074-4787},
 N.~Lurkin$\,${\footnotemark[6]}\orcid{0000-0002-9440-5927},
 F.~Newson,
 C.~Parkinson\orcid{0000-0003-0344-7361},
 A.~Romano\orcid{0000-0003-1779-9122}, 
 J.~Sanders\orcid{0000-0003-1014-094X},
 A.~Sergi$\,${\footnotemark[30]}\orcid{0000-0001-9495-6115},
 A.~Sturgess\orcid{0000-0002-8104-5571},
 J.~Swallow$\,${\footnotemark[7]}\orcid{0000-0002-1521-0911},
 A.~Tomczak\orcid{0000-0001-5635-3567}
\vspace{0.5cm}

{\bf School of Physics, University of Bristol, Bristol, United Kingdom}\\
 H.~Heath\orcid{0000-0001-6576-9740},
 R.~Page,
 S.~Trilov\orcid{0000-0003-0267-6402}
\vspace{0.5cm}

{\bf School of Physics and Astronomy, University of Glasgow, Glasgow, United Kingdom}\\
 B.~Angelucci,
 D.~Britton\orcid{0000-0001-9998-4342},
 C.~Graham\orcid{0000-0001-9121-460X},
 D.~Protopopescu\orcid{0000-0002-3964-3930}
\vspace{0.5cm}

{\bf Faculty of Science and Technology, University of Lancaster, Lancaster, United Kingdom}\\
 J.~Carmignani$\,${\footnotemark[31]}\orcid{0000-0002-1705-1061},
 J. B.~Dainton,
 R. W. L.~Jones\orcid{0000-0002-6427-3513},
 G.~Ruggiero$\,${\footnotemark[32]}\orcid{0000-0001-6605-4739}
\vspace{0.5cm}

{\bf School of Physical Sciences, University of Liverpool, Liverpool, United Kingdom}\\
 L.~Fulton,
 D.~Hutchcroft\orcid{0000-0002-4174-6509},
 E.~Maurice$\,${\footnotemark[33]}\orcid{0000-0002-7366-4364},
 B.~Wrona\orcid{0000-0002-1555-0262}
\vspace{0.5cm}

{\bf Physics and Astronomy Department, George Mason University, Fairfax, Virginia, USA}\\
 A.~Conovaloff,
 P.~Cooper,
 D.~Coward$\,${\footnotemark[34]}\orcid{0000-0001-7588-1779},
 P.~Rubin\orcid{0000-0001-6678-4985}
\vspace{0.5cm}

{\bf Authors affiliated with an Institute or an international laboratory covered by a cooperation agreement with CERN}\\
 A.~Baeva,
 D.~Baigarashev$\,${\footnotemark[35]}\orcid{0000-0001-6101-317X},
 D.~Emelyanov,
 T.~Enik\orcid{0000-0002-2761-9730},
 V.~Falaleev$\,${\footnotemark[4]}\orcid{0000-0003-3150-2196}, 
 S.~Fedotov,
 K.~Gorshanov\orcid{0000-0001-7912-5962},
 E.~Gushchin\orcid{0000-0001-8857-1665},
 V.~Kekelidze\orcid{0000-0001-8122-5065},
 D.~Kereibay, 
 S.~Kholodenko$\,${\footnotemark[27]}\orcid{0000-0002-0260-6570},
 A.~Khotyantsev,
 A.~Korotkova,
 Y.~Kudenko\orcid{0000-0003-3204-9426},
 V.~Kurochka, 
 V.~Kurshetsov\orcid{0000-0003-0174-7336},
 L.~Litov$\,${\footnotemark[14]}\orcid{0000-0002-8511-6883},
 D.~Madigozhin$\,$\renewcommand{\thefootnote}{\fnsymbol{footnote}}\footnotemark[1]\renewcommand{\thefootnote}{\arabic{footnote}}\orcid{0000-0001-8524-3455},
 M.~Medvedeva,
 A.~Mefodev, 
 M.~Misheva$\,${\footnotemark[36]},
 N.~Molokanova,
 S.~Movchan,
 V.~Obraztsov\orcid{0000-0002-0994-3641},
 A.~Okhotnikov\orcid{0000-0003-1404-3522}, 
 A.~Ostankov$\,$\renewcommand{\thefootnote}{\fnsymbol{footnote}}\footnotemark[2]\renewcommand{\thefootnote}{\arabic{footnote}},
 I.~Polenkevich,
 Yu.~Potrebenikov\orcid{0000-0003-1437-4129},
 A.~Sadovskiy\orcid{0000-0002-4448-6845},
 V.~Semenov$\,$\renewcommand{\thefootnote}{\fnsymbol{footnote}}\footnotemark[2]\renewcommand{\thefootnote}{\arabic{footnote}}, 
 S.~Shkarovskiy,
 V.~Sugonyaev\orcid{0000-0003-4449-9993},
 O.~Yushchenko\orcid{0000-0003-4236-5115},
 A.~Zinchenko$\,$\renewcommand{\thefootnote}{\fnsymbol{footnote}}\footnotemark[2]\renewcommand{\thefootnote}{\arabic{footnote}}
\vspace{0.5cm}

\end{raggedright}

%
%

\setcounter{footnote}{0}
\newlength{\basefootnotesep}
\setlength{\basefootnotesep}{\footnotesep}

\renewcommand{\thefootnote}{\fnsymbol{footnote}}
\noindent
$^{\footnotemark[1]}${Corresponding authors: F.~Brizioli, D.~Madigozhin, \\
email: francesco.brizioli@cern.ch, dmitry.madigozhin@cern.ch}\\
$^{\footnotemark[2]}${Deceased}\\
\renewcommand{\thefootnote}{\arabic{footnote}}
$^{1}${Present address: Ecole Polytechnique F\'ed\'erale Lausanne, CH-1015 Lausanne, Switzerland} \\
$^{2}${Present address: Syracuse University, Syracuse, NY 13244, USA} \\
$^{3}${Present address: Faculty of Science and Technology, University of Lancaster, Lancaster, LA1 4YW, UK} \\
$^{4}${Present address: INFN, Sezione di Perugia, I-06100 Perugia, Italy} \\
$^{5}${Also at TRIUMF, Vancouver, British Columbia, V6T 2A3, Canada} \\
$^{6}${Present address: Universit\'e Catholique de Louvain, B-1348 Louvain-La-Neuve, Belgium} \\
$^{7}${Present address: CERN, European Organization for Nuclear Research, CH-1211 Geneva 23, Switzerland} \\
$^{8}${Present address: Institut f\"ur Kernphysik and Helmholtz Institute Mainz, Universit\"at Mainz, Mainz, D-55099, Germany} \\
$^{9}${Present address: Universit\"at W\"urzburg, D-97070 W\"urzburg, Germany} \\
$^{10}${Present address: European XFEL GmbH, D-22869 Schenefeld, Germany} \\
$^{11}${Present address: School of Physics and Astronomy, University of Glasgow, Glasgow, G12 8QQ, UK} \\
$^{12}${Present address: Institut f\"ur Physik and PRISMA Cluster of Excellence, Universit\"at Mainz, D-55099 Mainz, Germany} \\
$^{13}${Also at Dipartimento di Scienze Fisiche, Informatiche e Matematiche, Universit\`a di Modena e Reggio Emilia, I-41125 Modena, Italy} \\
$^{14}${Also at Faculty of Physics, University of Sofia, BG-1164 Sofia, Bulgaria} \\
$^{15}${Present address: Scuola Superiore Meridionale e INFN, Sezione di Napoli, I-80138 Napoli, Italy} \\
$^{16}${Present address: Instituto de F\'isica, Universidad Aut\'onoma de San Luis Potos\'i, 78240 San Luis Potos\'i, Mexico} \\
$^{17}${Present address: Institut am Fachbereich Informatik und Mathematik, Goethe Universit\"at, D-60323 Frankfurt am Main, Germany} \\
$^{18}${Present address: INFN, Sezione di Roma I, I-00185 Roma, Italy} \\
$^{19}${Also at Department of Industrial Engineering, University of Roma Tor Vergata, I-00173 Roma, Italy} \\
$^{20}${Also at Department of Electronic Engineering, University of Roma Tor Vergata, I-00173 Roma, Italy} \\
$^{21}${Also at Universit\`a degli Studi del Piemonte Orientale, I-13100 Vercelli, Italy} \\
$^{22}${Also at Universidad de Guanajuato, 36000 Guanajuato, Mexico} \\
$^{23}${Present address: Charles University, 116 36 Prague 1, Czech Republic} \\
$^{24}${Present address: DESY, D-15738 Zeuthen, Germany} \\
$^{25}${Present address: Max-Planck-Institut f\"ur Physik (Werner-Heisenberg-Institut), M\"unchen, D-80805, Germany} \\
$^{26}${Present address: Brookhaven National Laboratory, Upton, NY 11973, USA} \\
$^{27}${Present address: INFN, Sezione di Pisa, I-56100 Pisa, Italy} \\
$^{28}${Present address: Department of Physics, University of Warwick, Coventry, CV4 7AL, UK} \\
$^{29}${Present address: Center for theoretical neuroscience, Columbia University, New York, NY 10027, USA} \\
$^{30}${Present address: Dipartimento di Fisica dell'Universit\`a e INFN, Sezione di Genova, I-16146 Genova, Italy} \\
$^{31}${Present address: School of Physical Sciences, University of Liverpool, Liverpool, L69 7ZE, UK} \\
$^{32}${Present address: Dipartimento di Fisica e Astronomia dell'Universit\`a e INFN, Sezione di Firenze, I-50019 Sesto Fiorentino, Italy} \\
$^{33}${Present address: Laboratoire Leprince Ringuet, F-91120 Palaiseau, France} \\
$^{34}${Also at SLAC National Accelerator Laboratory, Stanford University, Menlo Park, CA 94025, USA} \\
$^{35}${Also at L.N. Gumilyov Eurasian National University, 010000 Nur-Sultan, Kazakhstan} \\
$^{36}${Present address: Institute of Nuclear Research and Nuclear Energy of Bulgarian Academy of Science (INRNE-BAS), BG-1784 Sofia, Bulgaria} \\


\newpage
\clearpage

\begin{thebibliography}{99}

\bibitem{ChPT1}
S. Weinberg,
\emph{Phenomenological Lagrangians},
\href{https://doi.org/10.1016/0378-4371(79)90223-1}
{Physica A: Statistical Mechanics and its Applications \textbf{96} (1979) 327}.

\bibitem{ChPT2}
J. Gasser, H. Leutwyler,
\emph{Chiral perturbation theory to one loop},
\href{https://doi.org/10.1016/0003-4916(84)90242-2}
{Annals of Physics \textbf{158} (1984) 142}.

\bibitem{ChPT3}
J. Gasser, H. Leutwyler,
\emph{Chiral perturbation theory: Expansions in the mass of the strange quark},
\href{https://doi.org/10.1016/0550-3213(85)90492-4}
{Nucl. Phys. B \textbf{250} (1985) 465}.

\bibitem{Bijnens:1992en}
J. Bijnens, G. Ecker, J. Gasser,
\emph{Radiative semileptonic kaon decays},
\href{https://doi.org/10.1016/0550-3213(93)90259-R}
{Nucl. Phys. B \textbf{396} (1993) 81}.

\bibitem{Braguta_PhysRevD.65.054038}
 V. V. Braguta, A. A. Likhoded, A. E. Chalov,
 \emph{T-odd correlation in the ${K}_{l3\ensuremath{\gamma}}$ decay},
\href{https://doi.org/10.1103/PhysRevD.65.054038}
{Phys. Rev. D \textbf{65} (2002) 054038}.

\bibitem{Kubis:2006nh}
B. Kubis, E. H. Muller, J. Gasser, M. Schmid,
 \emph{Aspects of radiative $K^+_{e3}$ decays},
\href{https://doi.org/10.1140/epjc/s10052-007-0215-9}
{Eur. Phys. J. C \textbf{50} (2007) 557}.

\bibitem{Khriplovich:2010rz}
 I. B. Khriplovich, A. S. Rudenko,
 \emph{$K^+_{l3\gamma}$ decays revisited: branching ratios and T-odd momenta correlations},
\href{https://doi.org/10.1134/S1063778811080102}
{Phys. Atom. Nucl. \textbf{74} (2011) 1214}.

\bibitem{Akimenko:2007zz}
S. A. Akimenko et al. (ISTRA+ Collaboration),
\emph{Study of $K^- \rightarrow \pi^{0} e^{-} \Bar{\nu_e} \gamma $ decay with ISTRA+ setup},
\href{https://doi.org/10.1134/S1063778807040114}
{Phys. Atom. Nucl. \textbf{70} (2007) 702}.

\bibitem{Polyarush:2020ocu}
 A. Y. Polyarush et al. (OKA Collaboration),
\emph{Study of $K^+ \rightarrow \pi^{0} e^{+} \nu \gamma $ decay with OKA setup},
\href{https://doi.org/10.1140/epjc/s10052-021-08895-2}
{Eur. Phys. J. C \textbf{81} (2021) 161}.

\bibitem{pdg} 
 R. L. Workman et al. (Particle Data Group),
\emph{Review of Particle Physics},
\href{https://doi.org/10.1093/ptep/ptac097}
{Prog. Theor. Exp. Phys. \textbf{2022} (2022) 083C01}.
 
\bibitem{Muller:2006gu}
E.  H.  Muller,  B.  Kubis, Ulf-G.  Meissner,
\emph{T-odd correlations in radiative $K^+_{l3}$ decays and chiral perturbation theory},
\href{https://doi.org/10.1140/epjc/s10052-006-0033-5}
{Eur. Phys. J. C \textbf{48} (2006) 427}.

\bibitem{Braguta_beyond}
  V. V. Braguta, A. A. Likhoded, A. E. Chalov,
\emph{$T$-odd correlation in the ${K}^{+}\ensuremath{\rightarrow}\ensuremath{\pi}l\ensuremath{\nu}\ensuremath{\gamma}$ decays beyond the Standard Model},
\href{https://doi.org/10.1103/PhysRevD.68.094008}
{Phys. Rev. D \textbf{68} (2003) 094008}.

\bibitem{OKA_T}
 A. Y. Polyarush et al. (OKA Collaboration),
\emph{Measurement of the T-odd correlation in the $K^+ \rightarrow \pi^{0} e^{+} \nu_e \gamma $ radiative decay at the OKA setup},
\href{https://doi.org/10.1134/S0021364022602184}
{JETP Lett. \textbf{116} (2022) 608}.
 
\bibitem{pinunu}
E.  Cortina Gil et al. (NA62 Collaboration),
\emph{Measurement of the very rare K$^{+}$\textrightarrow{}$ {\pi}^{+}\nu \overline{\nu} $ decay},
\href{https://doi.org/10.1007/JHEP06(2021)093}
{JHEP \textbf{06} (2021) 093}.
 
\bibitem{detector}
E.  Cortina Gil et al. (NA62 Collaboration),
\emph{The beam and detector of the NA62 experiment at CERN},
\href{https://doi.org/10.1088/1748-0221/12/05/P05025}
{JINST \textbf{12} (2017) P05025}.

\bibitem{NA62:2020pwi}
E.  Cortina Gil et al. (NA62 Collaboration),
\emph{Search for $\pi^0$ decays to invisible particles},
\href{https://doi.org/10.1007/JHEP02(2021)201}
{JHEP \textbf{02} (2021) 201}.
 
\bibitem{readout}
R. Ammendola et al.,
\emph{The integrated low-level trigger and readout system of the CERN NA62 experiment},
\href{https://doi.org/10.1016/j.nima.2019.03.012}
{Nucl. Instrum. Meth. A \textbf{929} (2019) 1}.

\bibitem{trigger}
E.  Cortina Gil et al. (NA62 Collaboration),
\emph{Performance of the NA62 trigger system},
\href{https://doi.org/10.1007/JHEP03(2023)122}
{JHEP \textbf{03} (2023) 122}.

\bibitem{geant4}
J. Allison et al.,
\emph{Recent developments in Geant4},
\href{https://doi.org/10.1016/j.nima.2016.06.125}
{Nucl. Instrum. Meth. A \textbf{835} (2016) 186}.

 \bibitem{Gatti:2005kw}
C. Gatti,
\emph{Monte Carlo simulation for radiative kaon decays},
\href{https://doi.org/10.1140/epjc/s2005-02435-2}
{Eur. Phys. J. C \textbf{45} (2006) 417}.

 \bibitem{Barberio:1993qi}
 E. Barberio, Z. Was,
\emph{PHOTOS - a universal Monte Carlo for QED radiative corrections: version 2.0},
\href{https://doi.org/10.1016/0010-4655(94)90074-4}
{Computer Physics Communications \textbf{79} (1994) 291}.

\end{thebibliography}
\end{document}